\documentclass[letterpaper, 10 pt, conference]{ieeeconf}  

\IEEEoverridecommandlockouts                              

\usepackage{graphicx}
\usepackage{tabularx}
\usepackage{amsmath}
\usepackage{comment}
\title{\LARGE \bf
Variation is the Norm: Brain State Dynamics Evoked By Emotional Video Clips
}

\author{Ashutosh Singh$^{1,*}$, Christiana Westlin$^{2,*}$, Hedwig Eisenbarth$^{3}$, Elizabeth A. Reynolds Losin$^{4}$,\\
Jessica R. Andrews-Hanna$^{5,6}$, Tor D. Wager$^{7}$, Ajay B. Satpute$^{2}$,\\
Lisa Feldman Barrett$^{2,8,9}$, Dana H. Brooks$^{1}$, Deniz Erdogmus$^{1}$

\thanks{This work was supported by NSF (1947972). Data collection was supported by NIH R01 MH076136 and R01DA035484 (T.D.W).}
\thanks{$^*$Indicates shared first authorship. Correspondence should be addressed to Ashutosh Singh (singh.ashu@northeastern.edu).}
\thanks{$^{1}$Department of Electrical and Computer Engineering, College of Engineering, Northeastern University, Boston, MA, USA.}%
\thanks{$^{2}$Department of Psychology, College of Science, Northeastern University, Boston, MA, USA.} 
\thanks{$^{3} $School of Psychology, Victoria University of Wellington, Wellington, New Zealand}
\thanks{$^{4}$Department of Psychology, University of Miami, Miami, FL, USA}
\thanks{$^{5}$Department of Psychology, University of Arizona, Tucson, AZ, USA}
\thanks{$^{6}$Cognitive Science, University of Arizona, Tucson, AZ, USA}
\thanks{$^{7}$Department of Psychological and Brain Sciences, Dartmouth College, Hanover, NH, USA}
\thanks{$^{8}$Department of Psychiatry, Massachusetts General Hospital, Boston, MA, USA}
\thanks{$^{9}$Athinoula A. Martinos Center for Biomedical Imaging, Massachusetts General Hospital, Boston, MA, USA}%
}

\begin{document}
\maketitle
\thispagestyle{empty}
\pagestyle{empty}
\begin{abstract}
For the last several decades, emotion research has attempted to identify a ``biomarker" or consistent pattern  of brain activity to characterize a single category of emotion (e.g., fear) that will remain consistent across all instances of that category, regardless of individual and context. In this study, we investigated variation rather than consistency during emotional experiences while people watched video clips chosen to evoke instances of specific emotion categories. Specifically, we developed a sequential probabilistic approach to model the temporal dynamics in a participant's brain activity during video viewing. We characterized brain states during these clips as distinct state occupancy periods between state transitions in blood oxygen level dependent (BOLD) signal patterns. We found substantial variation in the state occupancy probability distributions across individuals watching the same video, supporting the hypothesis that when it comes to the brain correlates of emotional experience, variation may indeed be the norm.

\end{abstract}

\section{INTRODUCTION}

Emotions play a fundamental role in people's everyday life and are critical to understanding health and behavior. Models to understand and predict brain correlates of human emotional experience have become increasingly prevalent. The majority of these modeling approaches search for consistencies across sampled instances to identify a ``biomarker" for each commonly used emotion category, such as ‘anger,’ ‘sadness,’ ‘fear,’ etc. Although individual studies frequently report that they identify such biomarker patterns, the reported patterns are inconsistent across studies (e.g., compare \cite{kragel2015multivariate, saarimaki2016discrete, wager2015bayesian}, see \cite{clark2017multivoxel, azari2020comparing} for a discussion). In their attempt to identify consistent brain patterns that are specific to a given emotion category, studies tend to sample stimuli that limit the amount of variation that can be observed, and even when there is an opportunity to observe variable patterns in the neural correlates for a given category, models do not take advantage of the variation that is available to be modeled (\cite{azari2020comparing}). In the present study, we explicitly focused on this variation and hypothesized that the brains of participants display variable brain state dynamics while viewing the same emotional video clip.

Functional magnetic resonance imaging (fMRI) measures neural activity using the blood oxygen level dependent (BOLD) signal to index local relative increases in blood oxygenation due to activation-based increases in blood flow that exceed the rate at which oxygen is consumed \cite{ogawa1990brain, fox1988nonoxidative}. Researchers commonly measure the magnitude of the BOLD response under certain task conditions, each labeled as an instance of a specific emotion category (e.g., brain activity while watching a movie that was chosen to evoke ``anger" is labeled as an anger trial). A recent paper from our group compared solutions from supervised classification using a priori commonly used emotion category labels to those from unsupervised clustering in which no labels were assigned to the data and found that the two approaches did not produce concordant results \cite{azari2020comparing}. These findings suggested that commonly used emotion category labels might not be the best way to characterize emotion-related structure in brain data during emotional experience. However that study, along with the majority of published studies on emotional experience, focused only on task-based activation and ignored the temporal dynamics of the brain. 

In the present study, we investigated variation in the temporal dynamics of the BOLD signal during instances of emotional experience as people viewed movie clips that were curated to evoke emotions. Markov models have been previously used to investigate the organization of brain dynamics in the absence of task conditions (i.e., resting state fMRI; \cite{vidaurre2017brain}) as well as during tasks such as processing narrative stimuli (\cite{baldassano2017discovering}). Studies using these models (e.g., \cite{baldassano2017discovering}) typically make a hard assumption of only allowing one step forward state transitions. We relaxed this assumption in the present study by allowing backward state transitions (i.e., participants could re-enter a brain state they had previously been in throughout a sequence of video clips). We also used a mixture model for the observations (i.e. the fMRI data) given the state, allowing the possibility that the likelihood of the observations was more complex than a single Gaussian. We applied these methods to investigate the hypothesis that different individuals would vary in brain state dynamics while watching the same emotional video clip.

\section{Methods}
\subsection{Data Overview}

Eighty five participants (43 female, mean age = 28.79 years; min= 18; max= 54) participated in this study. All participants provided informed consent in accordance with guidelines set by the Institutional Review Board of the University of Colorado Boulder. Seven participants were excluded due to incomplete scans, resulting in a total sample size of 78 (42 female, mean age = 28.73 years; min= 18; max= 54). Participants underwent fMRI scanning while viewing six videos labeled based on a normed sample as ``loss" (i.e., "sadness"), ``anger," or ``disgust" (two videos per category). Videos were between 60 and 120 seconds in length (mean length 106 seconds). Following each video clip, participants provided a valence rating (i.e., ``How negative or positive do you feel right now?") for a  duration of 90s followed by a jittered ITI between 120 and 150ms.

Magnetic Resonance Imaging (MRI) scans were acquired on a Siemens MAGNETOM Trio 3T system (Siemens Medical Solutions, Erlangen, Germany) using a 32-channel head coil and included a structural MEMPRAGE scan (TR = 2530ms, TE1 = 1.64ms, TE2 = 3.5ms, TE3 = 5.36ms, TE4 = 7.22ms, TE5 = 9.08ms, FA = 7\textdegree, FOV = 256mm, voxel size 1mm isotropic, 192 slices) and a functional MRI scan (TR = 1300ms, TE = 25ms, FA = 50\textdegree, voxel size 3.4mm isotropic, FOV = 220mm, 26 slices).

\subsection{Data Preprocessing}
Image preprocessing was performed using the CONN toolbox \cite{whitfield2012conn} and included segmentation of gray and white matter tissue, realignment, slice-timing correction, normalization to Montreal Neurological Institute (MNI) space, spatial smoothing (8 mm FWHM Gaussian filter), and high-pass filtering using a Fast Fourier Transformation with a cut-off frequency of 0.008 Hz. The following potential confounding effects were regressed out: noise components from white matter and cerebrospinal areas, estimated subject-motion parameters (realignment and scrubbing parameters; outliers were identified as acquisitions with global signal changes above 5 s.d. or framewise displacement above 0.9mm), and constant task effects. To reduce the dimensionality of the data for analysis, the images were parcellated using the Glasser atlas \cite{glasser2016multi} which reduced the dimension of the whole-brain data to 180 regions averaged across hemispheres. We also conducted all analyses using 360 regions (180 per hemisphere) and achieved similar results, so for simplicity we only report results from the 180 regions averaged across hemispheres.  

\subsection{Modeling}

\begin{figure}[ht]
      \centering
     
      \includegraphics[scale=0.331]{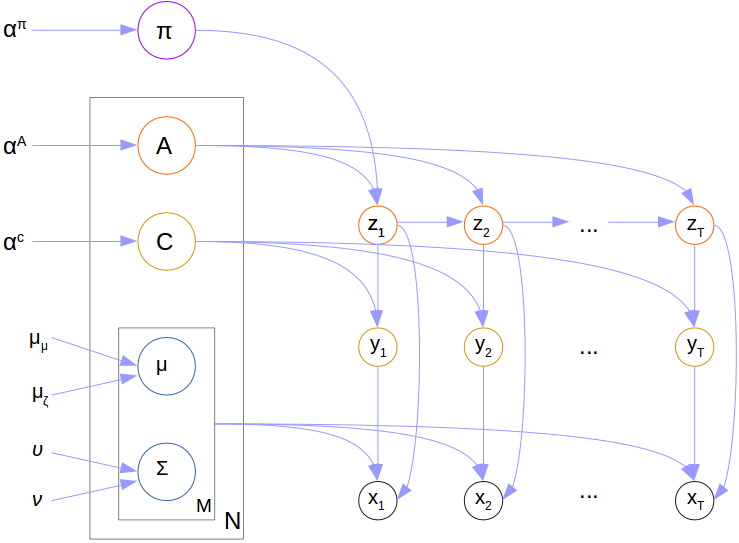}
      \caption{Block diagram for the Hidden Markov model with an emission probability density as a Gaussian Mixture model.}
      \label{fig:modelD}
   \end{figure}

Our probability model can be represented with the plate diagram shown in Figure \ref{fig:modelD}. The initial state prior is represented by $\pi$. The state at each time sample $t$ is indicated by the latent variable $z_t$, while $T$ represents the total number of fMRI volumes acquired during the presentation of a video clip. The transition probabilities between these states are represented by $A$ where each element $a_{ij}$ represents the probability of transitioning from state $i$ to state $j$ in the hidden Markov model (HMM). The model we use makes the assumption that the observation probability distribution (measurement x conditioned on state z) also known as emission probability density, takes the form of a Gaussian Mixture model (GMM). The latent variable $y_{t}$ indicates the mixture component label for this GMM, and the observed BOLD data is represented by $x_t$ (a 180x1 feature vector in this model). $C$ represents the prior for GMM components at each state; $c_{jm}$ is the probability of component $m$ at state $j$. $\mu$ and $\Sigma$ represent the mean vector and covariance matrix for each of the Gaussian distributions in the mixture. The model can therefore be specified by parameters $\theta = \{\pi ,A ,C ,\mu , \Sigma\}$. Equation \ref{equation: completedataliklihood} shows the complete data likelihood for this HMM-GMM model. 
\begin{equation}
\small
    p(x,z,y|\theta) = \pi_{z_1}
    \prod^{T-1}_{t=1}a_{z_t,z_{t+1}}
    c_{z_t,y_t} \mathcal{N}(x_t;\mu_{z_t,y_t},\Sigma_{z_t,y_t})
    \label{equation: completedataliklihood}
\end{equation}

We employed Dirichlet priors for each parameter set that represents a probability mass function ($\pi$, and each row of $A$ and $C$), with hyperparameters $\alpha_\pi$, $\alpha_A$ and $\alpha_C$, using the same hyperparameters for each row of $A$ and $C$. A normal distribution $\mathcal{N}$($\mu_\mu$,$\mu_\delta$) was used as a prior for the GMM mean vector and an inverse gamma distribution $G(u, v)$ was used as a prior for the covariance matrix. Here $\mu_{\mu}$ and  $\mu_\delta$ are the mean and covariance of the normal prior, and $u$ and $v$ are the shape and scale parameters of the inverse Gamma prior. We constrained our covariance matrix (for $x|y$) to be diagonal for ease of computation. The features represent average activity in distinct brain regions, so this assumption implies the model is committed to conditionally independent brain region responses for each GMM component conditioned on current brain state. The parameters $\theta$ are optimized according to the maximum likelihood parameter estimation principle, using a Baum-Welch algorithm \cite{rabiner1986introduction}. We used the following hyper-parameters for the model : $\alpha_{\pi}$ $=$ .5, $\alpha_{A}$ $=$ 1, $\alpha_C$ $=$ 1, $\mu_{\mu}$ $=$ 0, $\mu_{\delta}$ $=$ 0, $u$ $=$ -1.5 and $v$ $=$ 0.

\subsection{Analysis}

To select the number of state values, $N$, that the HMM can traverse through, and the number of mixture components, $M$, for each state value, we performed 6-fold cross-validation repeatedly (50 times with different 6-way splits) and calculated the average validation log-likelihood for each ($N$,$M$) pair. The pair with the highest validation log-likelihood averaged across the 50 repetitions was selected as the model order. For this model order, the remaining parameters were optimized to maximize the data likelihood over the entire dataset (all subjects, all video clips). This final training process assumed that the last brain state for one video clip could influence the first brain state of the following video clip. This process results in one model for all subjects and all video clips.

Using this model, we tested the hypothesis that individuals have different brain state sequences for a given video clip. We used the posterior probability distribution sequence of brain state given fMRI data (state occupancy probabilities) as a probabilistic surrogate for brain state sequences. Using a symmetric version of Kullback–Leibler (KL) divergence (i.e., Jensen–Shannon (JS) divergence, denoted by $D_{JS}$ in Equation~\ref{equation: kldivsym}), we compared the state occupancy probability distribution sequences for every pair of participants at every time step. JS divergence measures the dissimilarity between two probability mass functions, and takes normalized values in $[0,1]$.
Let $W = \frac{1}{2}(A+B)$ for two probability distributions $A$ and $B$. The KL and JS divergences are defined as: 

\begin{equation*}
D_{KL}(A \parallel W) = \sum_{x\in\mathcal{X}} A_x \log\left(\frac{A_x}{W_x}\right)
\end{equation*}
\begin{equation*}
D_{KL}(B \parallel W) = \sum_{x\in\mathcal{X}} B_x \log\left(\frac{B_x}{W_x}\right)
\end{equation*}
\begin{equation}
    {\rm D_{JS}}(A \parallel B)= \frac{1}{2}D_{KL}(A \parallel W)+\frac{1}{2}D_{KL}(B \parallel W).
    \label{equation: kldivsym}
\end{equation}

The inferred state occupancy probability of subject $s$ watching video clip $v$ at time $t$ is denoted by $P_{svt}$. Having calculated the sequence of $D_{JS}(P_{avt}\parallel P_{bvt})$ between subjects $a$ and $b$, we quantified time-averaged dissimilarity between the brain state sequences of these subjects using~(\ref{equation: kldivaveraged})
\begin{equation}
    D_{(a,b)v} = \frac{1}{T_v} \sum_{t=1}^{T_v} D_{JS}(P_{avt}\parallel P_{bvt})
    \label{equation: kldivaveraged}
\end{equation}
where $T_v$ represents the fMRI samples available for video clip $v$, which depends on its duration. This dissimilarity measure was used to perform hierarchical clustering with the single linkage algorithm, as described in \cite{mullner2011modern}. This clustering analysis allowed us to find groups of subjects with similar brain state sequences evoked by video clip $v$, if such groups exist.

\section{RESULTS}

\begin{figure}[ht]
      \centering
     
      \includegraphics[scale=0.4]{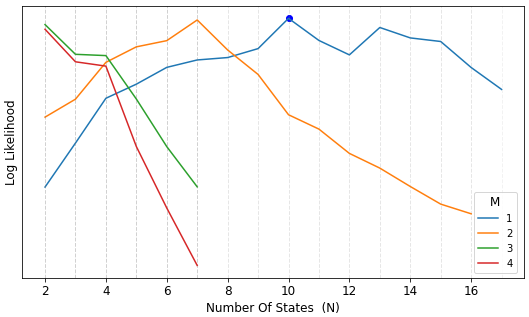}
      \caption{Results of 6-fold cross-validation for training the proposed model on all video clips across all participants. Each line represents a different number of mixture components (1 shown in blue, 2 in orange, 3  in green, 4  in red). M represents the number of mixture components, N represents the number of states. The blue dot for $M=1$ and $N=10$ indicates the model selected as best by this procedure.}
      \label{fig:kfold}
   \end{figure}
   
 Results from the 6-fold cross-validation model selection described above are shown in Figure \ref{fig:kfold}. The negative log likelihood was maximized with one mixture component ($M$ = 1) and ten states ($N$ = 10), and so we used these parameters in our model. We also evaluated results with the second best option, $M$ = 2 and $N$ = 7. The results from analyses using both parameter options yield similar results, and so for simplicity we only report results for $M$ = 1 and $N$ = 10.

\begin{figure}[ht]
      \centering
     
      \includegraphics[scale=0.29]{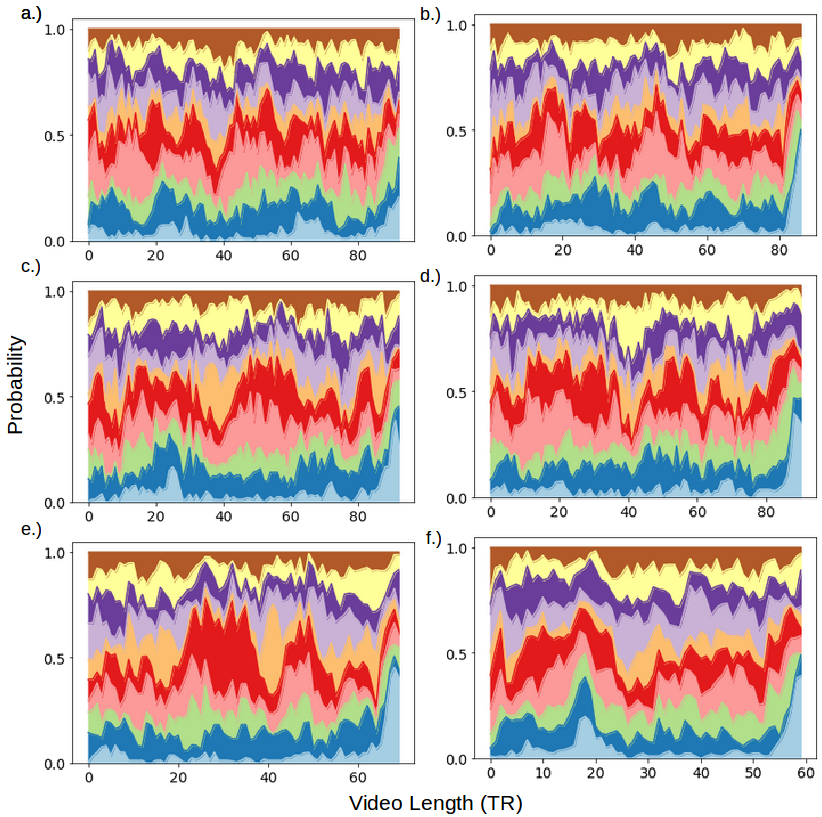}
      \caption{Probability of state activation averaged across participants for each clip at each time sample. Each plot represents a different video clip (graphs a-f represent video clips 1-6 respectively). Video clips 1 and 2 were categorized as loss, 3 and 4 as disgust, and 5 and 6 as anger. States 1-10 are represented by different colors from bottom (state 1) to top (state 10).) }
      \label{fig:avgstateactivation}
   \end{figure}

After running the model with 10 states and one mixture component, we examined the dynamic state occupancy probabilities for each video clip for each state. At every time instance we obtained a probability of observing each state (i.e., each time point had 10 associated probability values corresponding to a probability mass function). Figure \ref{fig:avgstateactivation} shows the state occupancy probability distributions for each clip averaged across participants as a function of time sample. The probability trajectories for these 10 states were distinct for each video, even when the videos were labeled with the same emotion category. Certain states had higher probability values for some videos than others (e.g., state 5 shows high probability during the middle of video clip 5, show in red in Fig. \ref{fig:avgstateactivation}e)), but these states did not have similar high probabilities across all clips.

\begin{figure}[ht]
      \centering
     
      \includegraphics[scale=0.36]{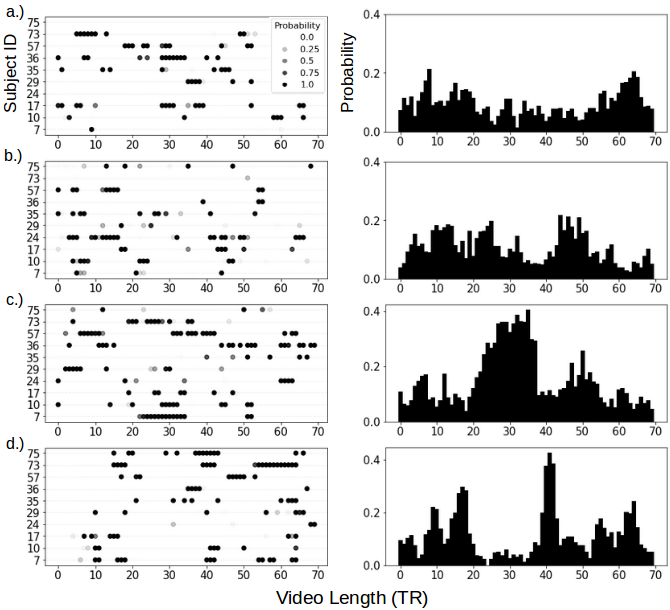}
      \caption{Occupancy probabilities for states 3, 4, 5 and 6 (shown in a-d respectively) across one representative clip (clip 5, anger). The left panel for each state depicts occupancy probabilities for ten selected participants. The opacity of the scatter plots is directly proportional to the probability values for the give state at the given TR, with darker regions indicating higher probabilities. The right panel depicts histograms of the average state occupancy probability across all participants for each of the representative states.}
      \label{fig:state1_individ}
   \end{figure}

We next investigated the state probability dynamics for each participant as they watched each clip rather than averaging across all participants. For example, the left panel  in  Figure \ref{fig:state1_individ} show occupancy profiles for four representative states (states three, four, five and six shown in a-d respectively) for one example clip (anger, clip five) across ten representative participants. The opacity of the points represents the probability value, with more opaque points reflecting higher values. Most participants had a high probability value for a single state at any given time point (i.e., a dominant state), rather than low values spread across all states. Participants also exhibited different temporal dynamics for a given state, transitioning in and out of the same state at different time points and spending variable amounts of time within a state, demonstrating variation in the occupancy probability dynamics at an individual participant level. This participant level variation in brain state sequences was not captured when averaging across all participants, as indicated in the right panel of Figure \ref{fig:state1_individ} .

\begin{figure}[ht]
      \centering
     
      \includegraphics[scale=0.5]{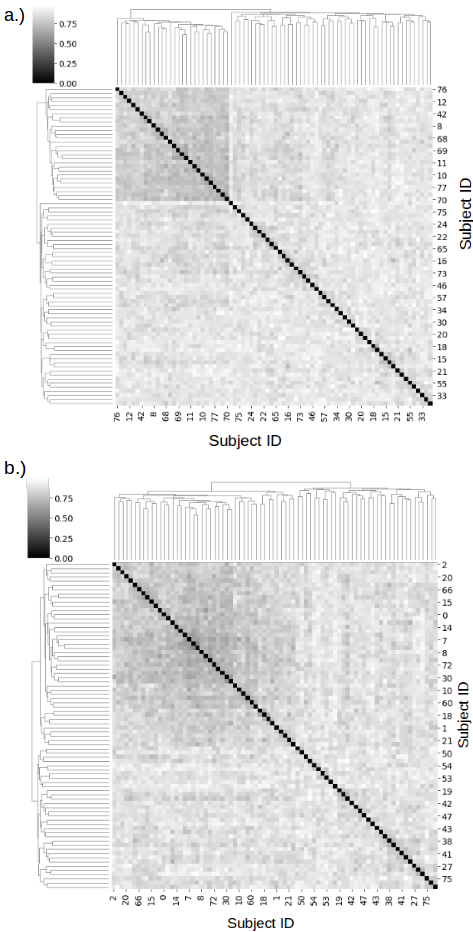}
      \caption{Subject-by-subject divergence matrices for two example clips (panels a and b reflect divergences for clips 4 and 5, respectively). Divergence is represented by $D_{(s_a,s_b)4}$ and $D_{(s_a,s_b)5}$ as defined by Equation 3. Subjects are ordered based on hierarchical clustering results. Darker regions indicate lower divergence values i.e.\ greater similarity.}
      \label{fig:kldiv}
   \end{figure}

Finally, we report results for symmetric KL divergence (i.e., JS divergence) to quantify the variation across participants in the state occupancy probability distributions for each clip. Figure \ref{fig:kldiv} a and b shows participant-by-participant divergence represented by Equation (3) for two representative video clips (video clips 4 and 5 (anger and disgust), respectively) for all subjects who watched each clip. These divergence matrices reveal high divergence of all participants from one another as they watched the same video clip (i.e., for a given video clip, the state occupancy probability distributions for each participant were highly dissimilar to all other participants). We performed hierarchical clustering on the participant-by-participant divergence matrices to investigate whether subgroups of participants were more similar to each other across video clips. Results revealed a subgroup of subjects who showed smaller divergence (i.e., higher similarity in state occupancy probability distributions) amongst themselves when compared to others. However, a subgroup of more similar participants did not exist for every video, and for videos where this subgroup did exist, the subgroup did not contain the same participants (e.g., subjects that comprise the darker, less divergent values in Figure \ref{fig:kldiv}a differ from those that comprise the darker, less divergent values in Figure \ref{fig:kldiv}b). The mean of these divergence distributions for videos 1-6 are 0.858, 0.856, 0.855, 0.852, 0.830 and 0.857, respectively, indicating high average divergence across all participants for all clips.

\section{DISCUSSION}

In the present study, we used a HMM-GMM model to investigate whether individuals vary in their brain state dynamics while watching the same set of videos that were curated to evoke emotional experiences. Our model revealed 10 common states across all video clips and participants, where each state represented a particular GMM model of the distribution of length 180 vectors of  regional averages of the observed BOLD data. The occupancy probabilities for each of these states were highly distributed when averaged across participants, such that there was some probability of being in each state at all time instances for every clip. This even spread of probabilities was likely caused by averaging across participants with variable occupancy probabilities, as was confirmed when we investigated the state occupancy probability distributions for each participant and video clip separately. We observed highly variable distributions across participants viewing the same video, such that participants transitioned in and out of the 10 states in a unique manner when viewing the same clip. We quantified the participant-by-participant variation in the state occupancy probability distributions for each video clip using JS divergence and observed high divergence across all participants for each clip. Our findings suggest that the brain state dynamics evoked by the same video clip, chosen to provoke a specific instance of an emotion category, were not consistent across participants, as is often assumed in emotion research, but were instead highly variable across participants for a given instance. We observed a subgroup of individuals that showed less variation within the subgroup (i.e., smaller divergence value) for some videos. However not all videos had a distinguishable subgroup, and for those videos that did, the subgroup was not consistently made up of the same participants. This may suggest that groups of individuals may indeed share similar brain state dynamics for a given instance of a given emotion, but these shared dynamics are not always present for a given instance, nor are they generally shared by the same group of subjects across instances. Future work is needed to investigate the anatomical basis of the reported states (e.g.\ which brain parcels are active for each state), and to determine the generalizability of the identified HMM states across other measurements  of emotional experience. More broadly our results suggest that researchers should consider adopting data-driven methods that do not rely on commonly used emotion category labels to investigate meaningful variation across individuals and instances of emotion.

\addtolength{\textheight}{-12cm}   


\bibliographystyle{IEEEtran}
\bibliography{ref}

\end{document}